\documentclass[12pt]{article}

\usepackage{amsfonts,amsmath,amssymb}
\usepackage{mathrsfs}
\usepackage{bm}
\usepackage{cite}

\usepackage[usenames]{color}

\allowdisplaybreaks[3]

\newcommand{\sfrac}[2]{{#1/#2}}

\newcommand{\correctedA}[1]{{#1}} 

\begin{document}

\begin{titlepage}

\begin{flushright}
KUNS-2321 \\
YITP-1121
\end{flushright}

\vspace{0.1cm}

\begin{center}
  {\Large
    On Non-Chiral Extension of Kerr/CFT
  }
\end{center}

\vspace{0.1cm}

\begin{center}

  Tatsuo {\sc Azeyanagi}$^{\dagger}$%
  \footnote{E-mail address : aze@gauge.scphys.kyoto-u.ac.jp},
  Noriaki {\sc Ogawa}$^{\ddagger}$%
  \footnote{E-mail address : noriaki@yukawa.kyoto-u.ac.jp}
  and
  Seiji {\sc Terashima}$^{\ddagger}$%
  \footnote{E-mail address : terasima@yukawa.kyoto-u.ac.jp}

\setcounter{footnote}{0}

\vspace{0.3cm}

${}^{\dagger}$
{\it Department of Physics, Kyoto University,\\
Kyoto 606-8502, Japan}\\

${}^{\ddagger}$
{\it Yukawa Institute for Theoretical Physics, Kyoto University,\\
     Kyoto 606-8502, Japan}
\end{center}

\vspace{1.5cm}

\begin{abstract}

We discuss possible non-chiral extension of the Kerr/CFT correspondence. 
We first consider the near horizon geometry of an extremal BTZ black hole 
and study the asymptotic symmetry. 
In order to define it properly, we introduce a regularization and 
show that the asymptotic symmetry becomes the desirable 
non-chiral Virasoro symmetry with the same central charges for both left and right sectors, 
which are independent of the regularization parameter.
We then investigate the non-chiral extension for general extremal black holes 
in the zero entropy limit. 
Since the same geometric structure as above emerges in this limit, 
we identify non-chiral Virasoro symmetry by a similar procedure.  
This observation 
supports the existence of a hidden non-chiral CFT$_2$ structure
with the same central charges for both left and right sectors dual to   
the rotating black holes. 
\end{abstract}
\end{titlepage}

\tableofcontents

\section{Introduction}
\label{sec:introduction}
Microscopic understanding of black holes is expected to provide us 
profound knowledge for the non-perturbative formulation of quantum gravity. 
Although string theory is one of the promising candidates for it, 
at present, only some restricted classes of black holes are microscopically 
dealt with by this theory \cite{Strominger:1996sh} 
(See also \cite{Peet:2000hn} for review).
Especially the detailed analysis of the Kerr black hole, 
which is astronomically the most familiar one, is not yet 
completed, though there are some 
proposals in this direction \cite{Horowitz:2007xq, Matsuo:2009sx}.

Parallel to the analysis based on string theory, another ``bottom up''
approach relying on asymptotic symmetry is widely carried out and 
is actively investigated recent years. 
From the viewpoint of gauge/gravity duality,  
the asymptotic symmetry of gravity side is identified with (a part of) 
the symmetry of the dual field theory living on the boundary.
One of the merits of this approach is that even when a black hole is not embedded in string theory yet, 
we can extract some information of the dual boundary theory,
in particular, when the theory is two dimensional and conformal.

Just after AdS/CFT correspondence is conjectured \cite{Maldacena:1997re}, 
the asymptotic symmetry approach is applied to Banados-Teitelboim-Zanelli (BTZ) black hole \cite{Banados:1992wn, Banados:1992gq, Strominger:1997eq},
a solution of three dimensional Einstein gravity with a  negative cosmological constant.
Since this black hole is asymptotically AdS$_3$,  the dual field theory is non-chiral CFT$_2$ and 
its central charges are already determined by the celebrated work by Brown and Henneaux \cite{Brown:1986nw}. 
Then by using Cardy's formula, the Bekenstein-Hawking entropy is microscopically reproduced. 
This result is remarkable itself and, at the same time, is very important for string theory.  
 This is because the BTZ black hole appears when we consider the near horizon limit of D1-D5-P black hole, 
 the first example string theory succeeded in deriving Bekenstein-Hawking entropy microscopically \cite{Strominger:1996sh}.
 
More recently, \correctedA{the asymptotic symmetry} approach is applied to extremal black holes 
whose near horizon geometry contains S$^1$ fibrated AdS$_2$ structure. 
The extremal Kerr black hole is included in this class and, since 
\correctedA{the} approach was first carried out to this black hole, 
this analysis is called the Kerr/CFT correspondence \cite{Guica:2008mu}.
Subsequently it was soon generalized to a large class of systems
\cite{Lu:2008jk,Azeyanagi:2008kb,Hartman:2008pb,Compere:2009dp,Azeyanagi:2009wf}
and they are also often called the Kerr/CFT correspondence, too.
In the Kerr/CFT, as a result of asymptotic symmetry approach, this class 
of extremal black holes are expected to be dual to chiral CFT$_2$ containing only single copy of Virasoro symmetry.       

This Kerr/CFT is very interesting, but,
to be precise, it is rather a suggestion for 
a possible pair of theories dual to each other
and there are many things to be understood to establish the Kerr/CFT as a concrete duality. 
One of the important problems is to derive the proper non-chiral
Virasoro symmetry with identical non-zero central charges, 
i.e.  $c_L=c_R$, from the gravity side. 
Although there are various evidence supporting the existence of the consistent non-chiral CFT$_2$ \correctedA{\cite{Bredberg:2009pv,Castro:2009jf,Hartman:2009nz,Castro:2010fd}},
unfortunately, no boundary condition allowing
such a  non-chiral Virasoro symmetry consistently has been obtained yet \cite{Matsuo:2009sj,Matsuo:2009pg,Rasmussen:2009ix,Rasmussen:2010sa}.
This would be because the non-chiral Virasoro symmetry is in some sense hidden \cite{Castro:2010fd}.

One possible path to understand non-chiral extensions of the Kerr/CFT is to consider 
the zero entropy extremal black holes 
\cite{Bardeen:1999px,Guica:2007gm, Balasubramanian:2007bs, Fareghbal:2008ar, Fareghbal:2008eh,
Nakayama:2008kg,Guica:2010ej,Azeyanagi:2010pw,Matsuo:2010ut,deBoer:2010ac}.
As shown in \cite{Azeyanagi:2010pw}, 
the extremal black holes with vanishing entropy and 
the regular AdS$_2$ structure\footnote{
An example is 5d Myers-Perry black holes with 
only one non-vanishing angular momentum.} 
will always have an emergent local AdS$_3$ structure.
For these black holes, therefore, a non-chiral CFT$_2$ is expected 
to reside as a dual field theory, and, following the work by
Brown and Henneaux \cite{Brown:1986nw},
we may be able to realize a non-chiral extension of the Kerr/CFT.

However, the ``local AdS$_3$'' appearing there has the same structure
as the near horizon extremal BTZ geometry,\footnote{We call the near horizon 
geometry of the extremal BTZ black hole in this manner.}
and the asymptotic geometry
is compactified on light-like circle and is different from 
the usual (global or Poincare) AdS$_3$ boundary where 
the asymptotic charges of \cite{Brown:1986nw} are defined.\footnote{
Furthermore, in the zero entropy limit, the radius of the circle becomes zero and thus the geometry
is singular, 
in the same way as the near horizon geometry of a massless BTZ black hole.
In this paper, we simply assume this does not 
cause any problem. We would like to return to this issue in future.}
Partly because of this, as we will see,  the asymptotic symmetry is not well defined 
for this geometry.

In this paper, we study the geometry slightly deformed 
as \correctedA{a} regularization of the light-like circle.
Indeed, we will show that the asymptotic symmetry 
becomes the non-chiral Virasoro symmetry with the same 
non-zero central charges for both sectors, $c_L=c_R$. 
Especially these charges are independent of the regularization parameter, 
justifying our regularization scheme. From the viewpoint of dual CFT$_2$, 
it is natural to expect that the central charges derived from the two geometries coincide,
since the Virasoro symmetries and central charges are ``local'' properties of the theory. 
Therefore, in this way, we can obtain a desirable result
and show that the Kerr/CFT can be understood as
the AdS$_3$/CFT$_2$ in the zero entropy limit.
The left-mover (Frolov-Thorne \cite{Frolov:1989jh}) temperature%
\footnote{
Below in this paper, ``temperature'' of given geometries
always stands for the Frolov-Thorne temperature, unless otherwise noted.
Notice that it is different from the Hawking temperature.
The Hawking temperature is always zero for extremal black holes.
}
of the system will be shown to be proportional to
the radius of the circle and, thus, vanishes in the limit, as expected.
Note that we here consider the Lorentzian version
of the AdS$_3$/CFT$_2$ where the temperature
of the system is 
proportional to the radius of the compactified circle \cite{Maldacena:1998bw}.
This rather strange duality is expected for
the Kerr/CFT in which the dual field theory side is a finite 
temperature system with Lorentzian signature.

We also show that one parameter family of limits leading to emergent local AdS$_3$ 
factor can be taken for the zero entropy extremal black holes.\footnote{
This can be possible probably because the geometry is singular.}
The parameter is a ratio of an infinitesimal parameter for the near horizon limit and that \correctedA{for the}
zero entropy limit. Especially, in \cite{Azeyanagi:2010pw}, the near horizon limit is taken first.
We then show that the geometry obtained is precisely the regularized one
we employed to derive the asymptotic symmetry properly,
and the ratio of the parameters for the two limits plays a role 
of the regularization term. 
Thus, the regularization can be naturally understood in this way, 
once we start with \correctedA{the} whole black hole geometry and then take these two limits carefully.%
\footnote{In \cite{Matsuo:2010ut}, a similar limit is introduced. Our result 
provides a natural interpretation to this limit and the regularization.}

Organization of this paper is as follows. In section~\ref{sec:AdS3}, 
we investigate a boundary condition and asymptotic symmetry for
the near horizon geometry of an extremal BTZ black hole.
We show that it contains two sets of Virasoro symmetries,
one of which corresponds to the chiral Virasoro symmetry of the Kerr/CFT for extremal black holes with non-zero entropy.
However, the other Virasoro symmetry is not centrally extended
by this naive prescription.
Then, to derive the asymptotic symmetry properly, we introduce a 
regularization to make the equal-time slice at the boundary space-like, 
and show that the desirable non-chiral Virasoro symmetry with 
the same central extensions for both left and right sector is realized.  
In section~\ref{sec:zeroentropy}, we show that a similar argument is applicable to 
the zero entropy extremal black holes. In section~\ref{sec:conclusion} we end up with 
conclusions and discussions. In the appendices, we summarize some 
results which might be useful for future analysis of the Kerr/CFT. 
In Appendix~\ref{sec:general_NHEK}, we explain that the similar regularization is 
not valid for the near horizon geometries of general (non-zero entropy) extremal black holes.
In Appendix~\ref{sec:mapping}, 
we provide a direct relation between the analysis of the 
asymptotic symmetries of
the AdS$_3$ and the near horizon extremal BTZ geometry.
In Appendix~\ref{sec:limits}, 
\correctedA{
we summarize various ways of taking the limits of zero entropy and near horizon
at the same time, by using the 5d extremal Myers-Perry black hole as a concrete example.
It proves that they form a
one parameter family and always lead to an emergent local AdS$_3$ structure.}
Especially, by considering the case in which the near horizon limit is taken faster than the zero entropy limit, 
we show that the regularization term is naturally introduced as a remnant of the whole black hole geometry.
In Appendix~\ref{sec:BTZ},
we also explain some relations between our results and the (holographic) renormalization group (RG) flow for
the BTZ black hole.

\section{Non-Chiral Kerr/CFT for Extremal BTZ Black Hole}
\label{sec:AdS3}

In this section, we deal with 
the near horizon geometry of an extremal BTZ black hole
and study the Kerr/CFT on it.
Since a BTZ black hole appears in the near horizon region of
the D1-D5-P black hole,
the analysis in this section can also be regarded as that for
this system. 

We start with a new boundary condition for this geometry,
giving a non-chiral extension of the Kerr/CFT there,
and explain only one of two Virasoro symmetries appearing as the 
asymptotic symmetry is centrally extended. We regard this is due to 
the light-like character of the equal-time surfaces at the boundary. 
We then introduce an appropriate regularization to make 
them space-like and show both two Virasoro symmetries are centrally extended.

The near horizon extremal BTZ geometry
is important not only in its
own right, but also in that
the same structure generally appears as a part of the near horizon
geometry for the zero entropy extremal black holes. 
The detailed analysis on this setup will be 
carried out in the next section.

\subsection{Boundary condition and asymptotic symmetry}
\label{sec:AdS3_naive_KerrCFT}

The near horizon extremal BTZ geometry is
written as
\begin{align}
  ds^2 =
  \bar{g}_{\mu\nu}dx^{\mu}dx^{\nu}
  &=
  \frac{L^2}{4}
  \left[
    -r^2dt^2+
    \frac{dr^2}{r^2}
    +(d\phi-rdt)^2
  \right]
\nonumber\\
  &=
  \frac{L^2}{4}
  \left(
    \frac{dr^2}{r^2}
    -2rdtd\phi
    +d\phi^2
  \right),
  \label{eq:NHEK_AdS3}
\end{align}
where 
$L$ is the AdS$_3$ radius and
the $\phi$-direction is orbifolded as
\begin{align}
  (t,\phi)\sim(t,\phi+2\pi\ell), 
\label{eq:phi_periodicity}
\end{align}
where $\ell$ is a constant which is determined by the mass of the BTZ black hole. 
\correctedA{
The horizon of the black hole corresponds to $r=0$,
and then the Bekenstein-Hawking entropy is calculated as}
\begin{align}
  S_{\mathit{BH}} = \frac{\frac{L}{2}\cdot 2\pi\ell}{4G_3} = \frac{\pi\ell L}{4G_3},
\label{eq:AdS3_Sbh}
\end{align}
where $G_3$ is the 3d Newton constant.

Before the detailed analysis of the asymptotic symmetry, 
here we give a comment on the orbifolding introduced in \eqref{eq:phi_periodicity}. 
This orbifolding is physically quite different from the more popular one
applied to the conventional Poincare AdS$_3$,
\begin{align}
  ds^2 &= L^2\Big(-\rho^2d\tau^2 + \frac{d\rho^2}{\rho^2} + \rho^2d\psi^2\Big),
  \label{eq:AdS3_conventional_AdS3}
\\
  &(\tau,\psi)\sim(\tau,\psi+2\pi\ell_{\psi}), \qquad (0<\ell_{\psi}<1).
  \label{eq:AdS3_psi_orbifold}
\end{align}
The latter just cuts the cylindrical boundary of
\eqref{eq:AdS3_conventional_AdS3} into a narrower cylinder.
On the other hand, the orbifold \eqref{eq:phi_periodicity}
generates a thermal state \cite{Maldacena:1998bw},
corresponding to the non-zero $S_{\mathit{BH}}$ \correctedA{as in} \eqref{eq:AdS3_Sbh}.
We will explain this point more in detail near the end of this section.
For more complete classification of various orbifoldings for AdS$_3$,
see \cite{Banados:1992gq}.

In order to investigate the asymptotic symmetry for
this background geometry \eqref{eq:NHEK_AdS3} with the orbifolding \eqref{eq:phi_periodicity},
we need to impose a proper boundary condition on the fluctuation of the metric.
One possible choice is
\begin{align}
  g_{\mu\nu} = \bar{g}_{\mu\nu} + h_{\mu\nu},
\qquad
  h_{\mu\nu}\sim
  \begin{pmatrix}
    r^2 & r^{-2} & 1\\
        & r^{-3} & r^{-1}\\
        &       & 1
  \end{pmatrix},
\end{align}
where the order of the coordinates in the matrix is set to $(t,r,\phi)$.
This can be regarded as 3d version of the boundary condition introduced in \cite{Guica:2008mu}.
Under this boundary condition, together with an energy constraint condition,
we can obtain a chiral Virasoro symmetry as the asymptotic symmetry group (ASG).
It is the most naive application of the Kerr/CFT to this system carried out in \cite{Azeyanagi:2008dk}.
In this paper, instead of this, we are interested in the case where
the ASG includes two sets of Virasoro symmetry.
We find that it is realized by the following boundary condition,
\begin{align}
  g_{\mu\nu} = \bar{g}_{\mu\nu} + h_{\mu\nu},
\qquad
  h_{\mu\nu}\sim
  \begin{pmatrix}
    1 & r^{-1} & 1\\
      & r^{-3} & r^{-1}\\
      &       & 1
  \end{pmatrix}.
  \label{eq:AdS3_KerrCFT_bc}
\end{align}
This boundary condition is special to the near horizon extremal BTZ geometry \eqref{eq:NHEK_AdS3},
where the $(t, t)$-component of the metric vanishes.%
\footnote{
The boundary condition \eqref{eq:AdS3_KerrCFT_bc} also works well
for the near horizon extremal BTZ geometry in the global coordinate, 
in which $\bar{g}_{tt}\sim 1$.
}
Actually, as is explained in Appendix~\ref{sec:mapping},
it is connected with a known boundary condition for the AdS$_3$
in the global or Poincare coordinate \cite{Porfyriadis:2010vg}.
Under \eqref{eq:AdS3_KerrCFT_bc},
the ASG is generated by
\begin{align}
  \xi =
    \big[f(t) + \mathcal{O}(1/r)\big]\partial_t
    + \big[-r(f'(t)+g'(\phi)) + \mathcal{O}(1)\big]\partial_r
    + \big[g(\phi) + \mathcal{O}(1/r)\big]\partial_\phi,
  \label{eq:AdS3_KerrCFT_ASG_form}
\end{align}
where $f(t)$ and $g(\phi)$ are respectively arbitrary functions of $t$ and $\phi$,
satisfying the periodicities imposed.
If we take the periodicity for $t$ as $t\sim t+\beta$ by hand, 
where $\beta$ is an arbitrary positive constant, 
the Fourier bases are $e^{-2\pi in\sfrac{t}{\beta}}$ and $e^{-in\sfrac{\phi}{\ell}}$ where 
$n$ is an integer. 
Especially, if we fix their normalizations as 
\begin{align}
  f_n(t) = -\frac{\beta}{2\pi}e^{-\frac{2\pi}{\beta}int},
\qquad
  g_n(\phi) = -\ell e^{-in\frac{\phi}{\ell}},
\label{eq:AdS3_naive_KerrCFT_Fourier_basis}
\end{align}
the corresponding bases for the ASG generators
\begin{subequations}
\begin{align}
  \xi^R_n &= -\frac{\beta}{2\pi}e^{-\frac{2\pi}{\beta}int}\partial_t 
             -inr e^{-\frac{2\pi}{\beta}int}\partial_r,
  \label{eq:AdS3_naive_KerrCFT_right_generators}\\
  \xi^L_n &= - \ell e^{-in\frac{\phi}{\ell}}\partial_{\phi}
             -inr e^{-in\frac{\phi}{\ell}}\partial_r,
  \label{eq:AdS3_naive_KerrCFT_left_generators}
\end{align}
\label{eq:AdS3_naive_KerrCFT_generators}%
\end{subequations}
composes two copies of Virasoro symmetry without central extension
\begin{align}
  [\xi_m^R, \xi_n^R] = -i(m-n)\xi_{m+n}^R,
\qquad
  [\xi_m^L, \xi_n^L] = -i(m-n)\xi_{m+n}^L,
\qquad
  [\xi_m^R, \xi_n^L] = 0.
\label{eq:AdS3_naive_KerrCFT_VirasoroAlgebra}
\end{align}
The Virasoro generators 
\eqref{eq:AdS3_naive_KerrCFT_left_generators}
have the same form as the chiral Virasoro generators in the Kerr/CFT,
and then this ASG with two Virasoro symmetries can be regarded as
a non-chiral extension of the Kerr/CFT.

The asymptotic charges corresponding to these generators 
also satisfy two copies of Virasoro symmetries and,  in 
this case, they can be centrally extended. 
To see the central extensions in detail and 
to confirm that the asymptotic charges are all consistent 
under the boundary condition \eqref{eq:AdS3_KerrCFT_bc},
we start with the definition of the asymptotic charges 
proposed by \cite{Barnich:2001jy}.
When the theory is $D$-dimensional Einstein gravity,%
\footnote{
Couplings with matter fields are also allowed,
and in general they may give some additional contributions to
\eqref{eq:asymptotic_charge_ktilde_formula}.
Here we assume that the matter fields are introduced such that 
they do not change the asymptotic symmetry consistently.  
In the usual Kerr/CFT, it is confirmed for various matter fields in \cite{Compere:2009dp}.
}
the asymptotic charge $Q_{\xi}=Q_{\xi}[h;\bar{g}]$
corresponding to an asymptotic symmetry generator $\xi$ and fluctuation $h_{\mu\nu}$,
defined on a $D$-dimensional background geometry $\bar{g}_{\mu\nu}$,
is given by
\begin{align}
  Q_{\xi} &= \int_{\partial\Sigma}\! k_{\xi}[h;\bar{g}],
\label{eq:asymptotic_charge_Q_formula}
\\
  k_{\xi}[h;\bar{g}]
  &= \tilde{k}_{\xi}^{\mu\nu}[h;\bar{g}]\,\frac{\epsilon_{\mu\nu\alpha_1\dots\alpha_{D-2}}}{(D-2)!}\,dx^{\alpha_1}\otimes\dots\otimes dx^{\alpha_{D-2}},
\label{eq:asymptotic_charge_k_formula}
\\
  \tilde{k}_{\xi}^{\mu\nu}[h;\bar{g}] &= -\frac{\sqrt{-\bar{g}}}{8\pi}\Big[
    \bar{D}^{[\nu}(h\xi^{\mu]})
    + \bar{D}_{\sigma}(h^{[\mu\sigma}\xi^{\nu]})
    + \bar{D}^{[\mu}(h^{\nu]\sigma}\xi_{\sigma})
\nonumber\\
    &\qquad\qquad\qquad\
    + \frac{3}{2}h\bar{D}^{[\mu}\xi^{\nu]}
    + \frac{3}{2}h^{\sigma[\mu}\bar{D}^{\nu]}\xi_{\sigma}
    + \frac{3}{2}h^{[\nu\sigma}\bar{D}_{\sigma}\xi^{\mu]}
  \Big],
\label{eq:asymptotic_charge_ktilde_formula}
\end{align}
and
the Poisson brackets of the asymptotic charges are given by,
\begin{align}
  \{Q_{\xi},Q_{\zeta}\}_{P.B.} = Q_{[\xi,\zeta]_\mathit{Lie}} + K_{\xi,\zeta},
\qquad
  K_{\xi,\zeta} = \int_{\partial\Sigma}\!k_{\zeta}[\pounds_{\xi}\bar{g};\bar{g}],
\label{eq:Poisson_bracket_formula}
\end{align}
where $K_{\xi,\zeta}$ is the central extension term.
Here the Greek indices are \correctedA{raised and lowered} by the background metric, 
$\bar{g} = \det(\bar{g}_{\mu\nu})$, $h = h_{\mu}{}^{\mu} = \bar{g}^{\mu\nu}h_{\mu\nu}$,
and $\Sigma$ is the $(D-1)$-dimensional equal-time hypersurface.
When we take $\Sigma$ at $t=\mathit{const}$ and the boundary at $r\to\infty$,
only the $(t,r)$ element of $\tilde{k}_{\xi}$ contributes to $Q_{\xi}$.

In the current case, $D=3$, by counting the order of $r$ in the $(t,r)$ element of each term of 
\eqref{eq:asymptotic_charge_ktilde_formula},
we can confirm that the asymptotic charges $Q^R_n$, $Q^L_n$ corresponding to
$\xi^R_n$, $\xi^L_n$ in \eqref{eq:AdS3_naive_KerrCFT_VirasoroAlgebra} respectively 
are all finite under the boundary condition \eqref{eq:AdS3_KerrCFT_bc}.
In this sense, we can say that \eqref{eq:AdS3_KerrCFT_bc} is a consistent boundary condition.%
\footnote{
Integrability of the charges is also necessary.
If we focus on small fluctuations around the background metric $\bar{g}$,
it can be easily shown in the same way as Appendix~B of \cite{Guica:2008mu}.
}
This is also special to the case of the extremal BTZ black hole,
and the situation is quite different for near horizon geometries for general extremal black holes, as  explained in Appendix~\ref{sec:general_NHEK}.

Explicit calculation of $K_{\xi,\zeta}$ for our asymptotic symmetry generators $\xi^R_n$ and $\xi^L_n$ yields 
Virasoro algebras with central extensions, 
\begin{subequations}
\begin{align}
[L_m^{R}, L_{n}^R] &= (m-n)L_{m+n}^R+\frac{c_R}{12}m(m^2-1)\delta_{m+n,0},  \\
[L_m^{L}, L_{n}^L] &= (m-n)L_{m+n}^L+\frac{c_L}{12}m(m^2-1)\delta_{m+n,0}, 
\end{align}
\label{eq:extended_Virasoros}%
\end{subequations}
where 
the Virasoro charges 
are defined as
\begin{align}
  L_n^{R,L} = Q_n^{R,L} + \delta_{n,0}\times(\mathit{const}),
\end{align}
and their quantum commutators are given by the classical Poisson brackets as
\begin{align}
  [\cdot,\cdot] =
  i\{\cdot,\cdot\}_{P.B.}.
\end{align}
The central charges are given by
\begin{align}
  c_{R,L} = 12i K_{\xi^{R,L}_{-n},\xi^{R,L}_{n}}\big|_{n^3},
\label{eq:central_charge_formula}
\end{align}
and the resulting values of the central charges are, respectively,
\begin{align}
  c_R = 0,
\qquad
  c_L = \frac{3L}{2G_3}.
\label{eq:AdS3_naive_KerrCFT_central_charges}
\end{align}

Therefore,
we have identified two Virasoro symmetries as the asymptotic symmetry, but
only one of the two is centrally extended. 
This situation is different from that in \cite{Brown:1986nw},
where 
both of the left and right Virasoro symmetries are centrally extended,
although the left central charge is the same.

There is a coordinate transformation between the near horizon extremal BTZ geometry and the conventional
form of AdS$_3$ used in \cite{Brown:1986nw},
and we can directly see the existence of an overlap region in the boundaries.  
Then the discrepancy above is somehow strange once we realize that 
the effect of the central charge is ``locally'' probed in CFT$_2$ by using,  
for example, the operator product expansion.

\subsection{Regularization for the light-like orbifolding}
\label{sec:AdS3_regularization}

In the last subsection, we took the ``equal-time hypersurface'' $\Sigma$
to be $t=\mathit{const}$.
Actually, in the current case, this naive prescription is problematic 
in that $\Sigma$ is not a space-like surface but a light-like one on the boundary.
To see this explicitly,
let us consider the metric induced 
on the conformal boundary of \eqref{eq:NHEK_AdS3}.
By taking $dr=0$ and $r\to \infty$, \eqref{eq:NHEK_AdS3} turns to be
\begin{align}
  ds^2
  &\stackrel{dr=0}{=} \frac{L^2}{4}( d\phi^2 - 2rdtd\phi)
\nonumber\\&
    \stackrel{r\to\infty}{\to} -\correctedA{\frac{L^2}{2}}rdtd\phi, 
\end{align}
Then the metric for the conformal boundary is 
\begin{align}
ds_{bdy}^2 =-dtd\phi. 
\label{eq:AdS3_NHEK_boundary_metric}
\end{align}
This metric vanishes for $dt=0$, and then the equal-time surface $\Sigma$
is light-like at the
boundary.
This suggests that our definition of the asymptotic charges
in the last subsection cannot be a proper one.

Now we would like to make $\Sigma$ space-like at the boundary.
It is, however, impossible under the ``light-like orbifolding''
\eqref{eq:phi_periodicity}.
Then 
we propose a little deformation of  
the geometry
by replacing \eqref{eq:phi_periodicity} with an orbifolding
along an ``infinitely boosted'' space-like direction $\phi'$.\footnote{
For related discussions, see \cite{Balasubramanian:2009bg, deBoer:2010ac}.
} 
More concretely, we define
\begin{align}
  t' = t + \alpha\phi,\quad \phi' = \phi,
\label{eq:AdS3_tprime_phiprime}
\end{align}
and replace \eqref{eq:phi_periodicity} by
\begin{align}
  (t',\phi')\sim (t',\phi'+2\pi\ell).
\label{eq:phiprime_periodicity}
\end{align}
Here $\alpha$ is a small positive constant introduced as a regularization
parameter.%
\footnote{
We can consider a coordinate transformation to general 
linear combinations of $t$ and $\phi$, but we confirmed that 
the final result is the same. } 
The new orbifolding \eqref{eq:phiprime_periodicity}
is different from the original periodicity \eqref{eq:phi_periodicity},
but they coincide in the $\alpha\to 0$ limit.
In the practical calculations of asymptotic charges and related quantities,
we first take the limit $r\to\infty$, and later take $\alpha\to 0$.
We exchanged the order of the limits here and 
it is the essence of the regularization trick.
Under \eqref{eq:phiprime_periodicity},
we regard $t'$ as ``time'' instead of $t$.
Because the ``equal-time surface'' $\Sigma'$ defined by $t'=\mathit{const}$
satisfies $dt=-\alpha d\phi$, it leads to
$ds^2_{\mathrm{bdy}}=\alpha d\phi^2 > 0$ and, 
therefore,  this surface is indeed space-like at the boundary.
\correctedA{As we will see in Appendix~\ref{sec:BTZ},
this procedure is naturally interpreted in terms of a remnant of the
original BTZ geometry before taking the near horizon limit.}

To respect the periodicity \eqref{eq:phiprime_periodicity},
the appropriate Fourier bases for $f(t)$ and $g(\phi)$
in \eqref{eq:AdS3_KerrCFT_ASG_form} are determined as
\begin{align}
  f_n(t) = -\alpha\ell e^{-in\frac{t}{\alpha\ell}},
\quad
  g_n(\phi) = -\ell e^{-in\frac{\phi}{\ell}},
\label{eq:AdS3_KerrCFT_Fourier_basis}
\end{align}
without introducing $\beta$ as in \eqref{eq:AdS3_naive_KerrCFT_Fourier_basis} by hand
--- $\beta$ is replaced by $2\pi\alpha\ell$ here.
The corresponding generators are
\begin{subequations}
\begin{align}
  \xi^R_n &= -(\alpha\ell \partial_{t} 
             +inr \partial_r) e^{-in\frac{t}{\alpha\ell}}
           = -(\alpha\ell \partial_{t'} 
             +inr \partial_r) e^{-in\sfrac{(\frac{t'}{\alpha}-\phi')}{\ell}},\\
  \xi^L_n &= -(\ell \partial_{\phi}
             +inr \partial_r) e^{-in\frac{\phi}{\ell}}
           = - (\alpha\ell \partial_{t'}
             + inr \partial_r
             + \ell \partial_{\phi'}) e^{-in\frac{\phi'}{\ell}},
\end{align}
\label{eq:AdS3_KerrCFT_generators}%
\end{subequations}
which satisfy Virasoro algebras in the same form as \eqref{eq:AdS3_naive_KerrCFT_VirasoroAlgebra}.
The background metric \eqref{eq:NHEK_AdS3} is also rewritten
by using $(t',\phi')$ as
\begin{align}
  ds^2 &= 
  \frac{L^2}{4}
  \left[
    \frac{dr^2}{r^2}
    -2rdt'd\phi'
    +(1+2\alpha r)d\phi'^2
  \right].
  \label{eq:AdS3regularized}
\end{align}
The corresponding asymptotic charges and their Poisson brackets are
defined and calculated by using
\eqref{eq:asymptotic_charge_Q_formula}, \eqref{eq:asymptotic_charge_k_formula},
\eqref{eq:asymptotic_charge_ktilde_formula} and \eqref{eq:Poisson_bracket_formula}
again, but $\Sigma$ is replaced by $\Sigma'$ there.
As a result, the contributing element of $\tilde{k}_{\xi}$ to the asymptotic charges 
is not $\tilde{k}_{\xi}^{tr}$, but $\tilde{k}_{\xi}^{t'r}$.%
\footnote{
Since $\tilde{k}_{\xi}^{t'r} = \tilde{k}_{\xi}^{tr} + \alpha \tilde{k}_{\xi}^{\phi r}$,
the difference between the results in \S\ref{sec:AdS3_naive_KerrCFT} and
those below comes from the contribution of the second term $\tilde{k}_{\xi}^{\phi r}$.
}
By a similar order counting of $r$ as in \S\ref{sec:AdS3_naive_KerrCFT},
$\tilde{k}_{\xi}^{t'r}$ is proved to include only finite terms for
all the ASG generators 
under the boundary condition \eqref{eq:AdS3_KerrCFT_bc}.
Therefore all the asymptotic charges are finite even when the regularization is introduced.

Furthermore, by using the formula for the central extension term \eqref{eq:Poisson_bracket_formula},
we obtain the finite value for $c_R$ as
\begin{align}
  c_R = \frac{3L}{2G_3}. 
\label{eq:AdS3_KerrCFT_right_central_charge}
\end{align}
This is exactly the value expected
and is the same as the right central charge derived by Brown and Henneaux. 
Remarkably, it does not depend on the value of $\alpha$ and 
then, in particular, it can be obtained in the limit $\alpha\to 0$.
This implies that our prescription works successfully as a regularization.
On the other hand, for $\xi^L_n$,
it leads to the same value for the central charge as \eqref{eq:AdS3_naive_KerrCFT_central_charges},
\begin{align}
  c_L = \frac{3L}{2G_3}.
\label{eq:AdS3_KerrCFT_left_central_charge}
\end{align}
These results \eqref{eq:AdS3_KerrCFT_right_central_charge} and \eqref{eq:AdS3_KerrCFT_left_central_charge}
are satisfying ones, in the viewpoints of Brown-Henneaux's analysis and
D1-D5 system in string theory.
Notice that, both of these central charges are independent of 
the periodicity of $\phi$, arbitrarily given by the orbifolding.
In fact this is always the case for the Kerr/CFT,%
\footnote{\label{fn:independence_of_periodicity}
This is very simply explained from
\eqref{eq:central_charge_formula}
and \eqref{eq:AdS3_KerrCFT_Fourier_basis}.
From \eqref{eq:central_charge_formula},
the central charge $c$ is bilinear to $\xi_n$'s and so
it gets a factor of $\ell^2$.
At the same time, since it is the coefficient of $n^3$ term,
the contributing terms include three derivatives of $f_n(t)$ or $g_n(\phi)$,
then they give a factor of $\sfrac{1}{\ell^3}$.
Furthermore, the boundary integral is carried out over $\phi\in[0,2\pi\ell)$
and it gives a factor $\ell$.
Therefore, in total, $c\sim\ell^{2-3+1}=\ell^0$,
which shows that the central charge does not depend on the
periodicity or orbifolding of $\phi$.
}
and it is consistent with the duality because
the central charges are local quantities in the dual CFT.

As for the right and left Frolov-Thorne (FT) temperatures 
$T^R_{\mathit{FT}}$, $T^L_{\mathit{FT}}$, we can determine them 
by employing the argument of \cite{Maldacena:1998bw} as follows. 
Let us consider a coordinate transformation 
\begin{align}
w_{+} = e^{\phi'},\quad w_{-} = -\frac{1}{2}\left(t'-\alpha \phi'+\frac{1}{r}\right), 
\quad y^2 =  \frac{1}{r}e^{\phi'}. 
\label{eq:mapping_w}
\end{align}
Then \eqref{eq:AdS3regularized} is rewritten in the form 
\begin{align}
ds^2 = L^2\frac{dw_{+}dw_{-}+dy^2}{y^2}.
\label{eq:AdS3_by_wy}
\end{align}
At the boundary, the relation between $w_{+}$ and $\phi'$ are similar to 
the one for (a half of) the Minkowski coordinate and 
the Rindler coordinate. Therefore, the left modes corresponding to 
$\phi'$ is in thermal state. Following \cite{Maldacena:1998bw}, 
its temperature is related to the periodicity as $w_{+}\sim e^{4\pi^2T^{L}_{\mathit{FT}}}w_{+}$.   
Now that $\phi'\sim\phi'+2\pi \ell$, the left FT temperature is  
\begin{align}
T^L_{\mathit{FT}} = \frac{\ell}{2\pi}.
\label{eq:AdS3_FT_temperatures_L}
\end{align}
On the other hand,  for $w_{-}$, since it is linearly related to $t'$ and $\phi'$ at the boundary, 
the corresponding right mode is not thermal and we then obtain 
\begin{align}
T^R_{\mathit{FT}}=0. 
\label{eq:AdS3_FT_temperatures_R} 
\end{align}

Combining the results \eqref{eq:AdS3_KerrCFT_right_central_charge}, \eqref{eq:AdS3_KerrCFT_left_central_charge}, 
\eqref{eq:AdS3_FT_temperatures_L} and \eqref{eq:AdS3_FT_temperatures_R},
we can calculate the entropy on the boundary CFT by \correctedA{Cardy's} formula
as
\begin{align}
  S_{\mathrm{Kerr/CFT}}
  = \frac{\pi^2}{3}c_RT^R_{\mathit{FT}} + \frac{\pi^2}{3}c_LT^L_{\mathit{FT}}
  = \frac{\pi\ell L}{4G_3},
\end{align}
which perfectly agrees with the Bekenstein-Hawking entropy \eqref{eq:AdS3_Sbh}.

The calculations in this subsection are similar to the ones 
in \S 5 of \cite{Matsuo:2010ut},
but the interpretation is different.
Here we emphasize again that our analysis depends on the structure of the near horizon extremal BTZ geometry.
It is then applicable to the general zero entropy extremal black holes investigated in \cite{Azeyanagi:2010pw},
as we will see in the next section. We note that it is, however, not so for general
non-zero entropy extremal black holes.     
We will comment on some issues appearing for the latter cases in
Appendix~\ref{sec:general_NHEK}.

\section{Zero Entropy Black Holes}
\label{sec:zeroentropy}
As recently shown in \cite{Azeyanagi:2010pw},
a local AdS$_3$ structure appears in 
the zero entropy limit of general extremal black holes. 
In this case, it has the same orbifolding structure as the
near horizon extremal BTZ geometry,
and the periodicity shrinks to zero, similarly to the near horizon geometry
of the massless BTZ black hole.
By using the 
argument in the previous section, we can identify 
two sets of Virasoro symmetries with non-zero central charges as the asymptotic symmetry. 
Note that, by adding appropriate charges if necessary, 
the zero entropy limit could be taken for any black hole geometry.

Let us consider the near horizon geometry of the general 
4d extremal black holes with $SL(2,R)\times U(1)$ symmetry 
(generalization to higher dimensional cases would not be difficult)
The metric is generally written as  \cite{Kunduri:2007vf}
\begin{align}
ds^2 = A(\theta)^2\left[ -r^2dt^2+\frac{dr^2}{r^2}+B(\theta)^2(d\phi-krdt)^2 
\right]+F(\theta)^2d\theta^2, 
\end{align}
where $A(\theta)$, $B(\theta)$, $F(\theta)$ are functions of $\theta$ determined
by solving equations of motion and $k$ is a constant. 
When the entropy is very small, the metric can be written in the form \cite{Azeyanagi:2010pw}
\footnote{
Here, the expression $A=o(B)$ means that $\lim \sfrac{A}{B}=0$.
}
\begin{align}
  ds^2=A(\theta)^2
   \left[
     -r^2dt^2 + \frac{dr^2}{r^2}
     + \bigg(1+\frac{\beta(\theta)}{k^2} + o\Big(\frac{1}{k^2}\Big)\bigg)(d\phi - rdt)^2
   \right]
   + F(\theta)^2d\theta^2,
   \label{eq:4d_AdS3_with_subleading}
\end{align}%
with a periodicity
\begin{align}
  \phi\sim\phi + \frac{2\pi}{k}.
  \label{eq:smallentropy_phi_periodicity}
\end{align}
The zero entropy limit corresponds to $k\to\infty$ here.
In the limit,
an AdS$_3$ factor emerges and
the metric goes to
\begin{align}
  ds^2&=A(\theta)^2
   \left[
     -r^2dt^2 + \frac{dr^2}{r^2}
     + (d\phi - rdt)^2
   \right]
   + F(\theta)^2d\theta^2.
   \label{eq:4d_AdS3}
\end{align}
Here the periodicity of $\phi$ is
\begin{align}
  \phi\sim \phi + 2\pi\delta,
  \label{eq:zeroentropy_phi_periodicity}
\end{align}
where $\delta=1/k$ is taken to be infinitesimal.%
\footnote{\label{fn:replace_periodicity}
Actually, we introduce a regularization here.
The $(\sfrac{1}{k})$-suppressed term in \eqref{eq:4d_AdS3_with_subleading}
vanishes in the $k\to\infty$ limit and we get \eqref{eq:4d_AdS3},
but at the same time 
the period $2\pi/k$ in \eqref{eq:smallentropy_phi_periodicity} also goes to zero,
making the geometry singular.
Our prescription to avoid this difficulty is as follows.
We formally regard $\delta$ as a small constant which is independent of $k$,
and we take the $k\to\infty$ limit.
It leaves us the metric \eqref{eq:4d_AdS3} with \eqref{eq:zeroentropy_phi_periodicity}.
After that, we take $\delta\to 0$ limit at last. 
}
\correctedA{The black hole horizon corresponds to $r=0$.}
The \correctedA{Bekenstein-Hawking} entropy vanishes in the $k\to\infty$ limit under \eqref{eq:zeroentropy_phi_periodicity},
but for a very large but finite $k$,
it is given by, \cite{Azeyanagi:2010pw}
\begin{align}
  S_{\mathit{BH}}
  = \frac{\pi}{2kG_4}\int\!\!d\theta\, A(\theta)F(\theta)
    + \mathcal{O}(\sfrac{1}{k^3}).
\label{eq:zeroentropy_Sbh}
\end{align}

For this geometry, we can put a boundary condition as
\begin{align}
  h_{\mu\nu}\sim
  \begin{pmatrix}
    1 & r^{-1} & 1 & 1 \\
      & r^{-3} & r^{-2} & r^{-1} \\
      &       & r^{-1} & 1 \\
      &       &   & 1
  \end{pmatrix},
  \label{eq:zeroentropy_KerrCFT_bc}
\end{align}
where the order of the coordinates is $(t,r,\theta,\phi)$.
The corresponding ASG is generated, similarly to \eqref{eq:AdS3_KerrCFT_ASG_form}, by
\begin{align}
  \xi =
    \big[f(t) + \mathcal{O}(1/r)\big]\partial_t
    + \big[-r(f'(t)+g'(\phi)) + \mathcal{O}(1)\big]\partial_r
    + \big[g(\phi) + \mathcal{O}(1/r)\big]\partial_\phi
    + \mathcal{O}(1/r)\partial_{\theta}.
  \label{eq:zeroentropy_KerrCFT_ASG_form}
\end{align}
In the following step, our prescription is almost the same as \S\ref{sec:AdS3_regularization}.
In an exactly similar way to \eqref{eq:AdS3_NHEK_boundary_metric}, 
the hypersurface defined by $t=\mathit{const}$ is light-like at the boundary,
and then it is not appropriate to define asymptotic charges on it.
Then we adopt new coordinates $(t', \phi')$ as
\eqref{eq:AdS3_tprime_phiprime} and obtain the ASG generators
\begin{subequations}
\begin{align}
  \xi^R_n &= -(\alpha\delta \partial_{t} 
             +inr\partial_r) e^{-in\frac{t}{\alpha\delta}}
           = -(\alpha\delta \partial_{t'} 
             +inr \partial_r) e^{-in(\frac{t'}{\alpha}-\phi')/\delta},\\
  \xi^L_n &= - (\delta \partial_{\phi}
             +inr \partial_r) e^{-in\frac{\phi}{\delta}}
           = - (\alpha\delta \partial_{t'}
             + inr \partial_r
             + \delta \partial_{\phi'})
             e^{-in\frac{\phi'}{\delta}},
\end{align}
\label{eq:zeroentropy_KerrCFT_generators}%
\end{subequations}
which are the same as \eqref{eq:AdS3_KerrCFT_generators},
with $\ell$ replaced by $\delta$.
The metric is written by using $(t',\phi')$ as
\begin{align}
  ds^2&=A(\theta)^2
   \left[
    \frac{dr^2}{r^2}
    -2rdt'd\phi'
    +(1+2\alpha r)d\phi'^2
   \right]
   + F(\theta)^2d\theta^2,  
  \label{eq:zeroentropy_regularized}
\end{align}
where the periodicity is 
\begin{align}
  (t',\phi')\sim (t',\phi' + 2\pi\delta).
  \label{eq:zeroentropy_regularized_periodicity}
\end{align}
As we will explain in Appendix \ref{sec:limits},
this metric \eqref{eq:zeroentropy_regularized}
and the periodicity \eqref{eq:zeroentropy_regularized_periodicity}
appear when we carefully take the near horizon and zero entropy limits simultaneously
for 5d extremal Myers-Perry black hole,  
\eqref{eq:limits_MP_NH_ZE3} and \eqref{eq:limits_MP_NH_ZE3_periodicity},
though the latter contains some extra structure because it is a higher dimensional geometry.
Under this identification, in particular,
\begin{align}
  C = 2\alpha,
\label{eq:zeroentropy_origin_of_regularization}
\end{align}
and the $\alpha\to 0$ limit corresponds to $C\to 0$ in Appendix \ref{sec:limits}.
Therefore, the regularization \eqref{eq:AdS3_tprime_phiprime}
is introduced naturally and automatically,
once we recall that the near horizon geometry \eqref{eq:4d_AdS3}
comes from the whole black hole geometry and consider an infinitesimal residue from it.

To confirm the consistency of the current boundary condition \eqref{eq:zeroentropy_KerrCFT_bc}, 
we checked by order counting that the asymptotic charges
corresponding to \eqref{eq:zeroentropy_KerrCFT_generators} are all finite.
The central charges $c_R$, $c_L$ can also be calculated in a similar way to
\S\ref{sec:AdS3_regularization} and we have
\begin{align}
  c_R = c_L = \frac{3}{G_4}\int\!\!d\theta\,A(\theta)F(\theta).
\end{align}
From the periodicity \eqref{eq:zeroentropy_phi_periodicity},
the Frolov-Thorne temperatures are
\begin{align}
  T_{FT}^R = 0,
\qquad
  T_{FT}^L = \frac{\delta}{2\pi} = \frac{1}{2\pi k},
\end{align}
in the same way as \S\ref{sec:AdS3_regularization}.
Then the entropy is calculated as
\begin{align}
  S_{\mathrm{Kerr/CFT}}
  &= \frac{\pi^2}{3}c_RT^R_{\mathit{FT}} + \frac{\pi^2}{3}c_LT^L_{\mathit{FT}}
\nonumber\\
  &= \frac{\pi}{2kG_4}\int\!\!d\theta\,A(\theta)F(\theta),
\end{align}
which reproduces the Bekenstein-Hawking entropy \eqref{eq:zeroentropy_Sbh},
in the leading order of $\sfrac{1}{k}$.

\section{Conclusions and Discussions}
\label{sec:conclusion}

In this paper, we first studied the asymptotic symmetry for 
the near horizon extremal BTZ geometry.
Under an appropriate boundary condition, it includes two Virasoro symmetries and can be regarded as a non-chiral
extension of the Kerr/CFT.
However, by a naive prescription,
only one of the two Virasoro symmetries is centrally extended.
We recognized it is due to the light-like character of the equal-time surface at the
boundary. Then, by introducing  an appropriate regularization to make the 
equal-time surface at the boundary space-like, we showed that
both two are centrally extended consistently. 

Since the same geometric structure as the near horizon extremal BTZ geometry emerges in 
the zero entropy limit for general extremal black holes, as discussed in \cite{Azeyanagi:2010pw}, 
we then applied the regularization to this setup and showed the existence of 
non-chiral Virasoro symmetries with central extensions. 
In this context, we also explained that our regularization scheme has a natural interpretation 
as a ratio of the infinitesimal parameter for the near horizon limit and the one for the zero entropy limit .
At the same time, because the $S^1$ fiber shrinks in the zero entropy limit,
we introduced another regularization for it.
It is a subtle prescription and we do not have a strong justification for it,
but the desirable results obtained from this prescription suggests the validity of it in some sense.

We focused on the 4d zero entropy extremal black holes in \S\ref{sec:zeroentropy} for simplicity,
but extension to higher-dimensional systems would not be difficult.
In higher-dimensional zero entropy extremal black holes,
there are more than one way to take the zero entropy limit
and obtain the local AdS$_3$ structure.
That is, one of the several rotating directions shrinks and becomes a part of
this geometric structure.
Generally speaking, we have $D-3$ choices at most for $D$-dimensional systems.
This is also consistent with the fact that 
there are several ways to realize the Kerr/CFT
in higher-dimensional extremal black holes \cite{Lu:2008jk,Azeyanagi:2008kb}.

Although the current analysis is restricted to zero entropy extremal black holes, 
we hope this result would be valuable for the extension of the Kerr/CFT 
beyond extremal black holes. 
Here we again notice that the argument in the text is special to 
the local AdS$_3$ structure
and a similar one is not applicable to the near horizon geometries for general extremal black holes. 
As summarized in Appendix~\ref{sec:general_NHEK}, when applied 
to these geometries, 
there are some unsolved problems. 
We think that the situation is the same for general warped 
AdS$_3$ geometries discussed in the context of topologically massive gravity with a negative cosmological constant 
in three dimensions \cite{Li:2008dq, Anninos:2008fx}.  

In this paper, we focused on the near horizon extremal BTZ geometry and higher-dimensional geometries including it, 
but the first analysis of the asymptotic symmetry on AdS$_3$
is carried out for the global AdS$_3$ by Brown and Henneaux. 
It is then valuable to comment on some relation between these two analysis beyond the correspondence of the central charges.
In Appendix~\ref{sec:mapping},  an explicit form of the map between these two coordinates is summarized 
and, by using this, we directly show the existence of some overlap of boundary regions for the two coordinates 
as well as  the correspondence of the boundary conditions and the asymptotic symmetry generators.
In Appendix~\ref{sec:BTZ},
some relations between our analysis in this paper and
the holographic RG flow of the Virasoro generators in the extremal BTZ black hole are summarized.
They might be useful for deeper understanding of the asymptotic symmetry of \correctedA{the BTZ black hole} and other geometries.

\section*{Acknowledgements}
T.~A.~and N.~O.~are supported by the Japan Society for the Promotion of Science (JSPS).
S.~T.~is partly supported by the Japan Ministry of Education, Culture, Sports, Science and Technology (MEXT). 
T.~A.~would like to thank people in the Particle, Field, String Theory Group and INT in University of Washington for hospitality, where a part of this work is carried out.
\correctedA{S.~T.~would like to thank George Moutsopoulos for valuable comments.}
This work is supported by the Grant-in-Aid for the Global COE program
``The Next Generation of Physics, Spun from Universality and Emergence''
from the MEXT. 
\vspace{10mm}

\appendix

\section{General Near Horizon Geometries}
\label{sec:general_NHEK}

In \S\ref{sec:AdS3} and \S\ref{sec:zeroentropy}
we derived non-zero left and right central charges
for the near horizon extremal BTZ geometry and the zero entropy extremal black holes,
in a consistent manner by introducing an appropriate regularization.
As emphasized there, we made use of some special properties of
the local AdS$_3$ structure.

Now we would like to consider near horizon geometries for general extremal black holes,
which have, for example in 4d, the form of
\begin{align}
ds^2 &= 
A(\theta)^2
\left[
-r^2dt^2+\frac{dr^2}{r^2}+B(\theta)^2(d\phi-krdt)^2
\right]+F(\theta)^2d\theta^2,
\label{eq:4d_general_NHEK}\\
&(t,\phi)\sim (t,\phi+2\pi).
\label{eq:4d_general_NHEK_naive_periodicity}
\end{align}
It looks very similar to the near horizon extremal BTZ geometry, but 
it proves to be quite difficult to introduce an appropriate regularization.
In this appendix, we naively employ a similar regularization scheme and
see that divergence of the asymptotic charges is inevitable in this case.
We restrict our analysis to 4d geometries,
but generalization to higher dimensions would not be difficult.

In the near horizon geometry \eqref{eq:4d_general_NHEK},
the \correctedA{black hole horizon corresponds to} $r=0$, and the Bekenstein-Hawking entropy is
\begin{align}
  S_{\mathit{BH}} = \frac{\pi}{2G_4}\int\!d\theta\,A(\theta)B(\theta)F(\theta).
\end{align}
In order to realize non-chiral Kerr/CFT for this geometry,
we need to find a consistent boundary condition
allowing two Virasoro symmetries as ASG
and then find a way to compute the central charges correctly,
if needed, by introducing some regularization.

The boundary condition \eqref{eq:zeroentropy_KerrCFT_bc}
we employed for the zero entropy case does not work for
\eqref{eq:4d_general_NHEK}, since
\begin{align}
  \pounds_{\xi^L_n}\bar{g}_{tt} = 2ine^{-in\phi}A(\theta)^2\big(1-k^2B(\theta)^2\big)r^2\quad (\gg 1).
\end{align}
It cannot be canceled by the contribution from subleading terms
which might be added to $\xi^L_n$,
and vanishes only when $B(\theta)=\sfrac{1}{k}$,
that is, when the geometry has a local AdS$_3$ structure.
For the general case, some boundary conditions have been proposed,
such as \cite{Matsuo:2009pg,Rasmussen:2010sa},
but it is fair to say that we do not know any satisfying one yet.
For example, it would be difficult to consistently remove the divergence
of all the asymptotic charges in \cite{Matsuo:2009pg},
and 
there is no hope to obtain a non-zero $c_R$ for \cite{Rasmussen:2010sa}.

Other attempts have been made to find a boundary condition
whose ASG includes only the ``right-handed'' Virasoro symmetry
\cite{Matsuo:2009sj,Rasmussen:2009ix}, which is claimed to 
stand for a different sector (``right-movers'') of the same dual field theory as the usual Kerr/CFT (``left-movers'').
There might be some way to impose some consistent boundary condition
but we do not pursue this possibility here. Below we just assume its existence 
and discuss the effect of the regularization.

Let us then assume that
ASG respecting such a boundary condition includes the two Virasoro algebras generated by
\begin{align}
  \xi^R_n = f_n(t)\partial_t -rf'_n(t)\partial_r,
\qquad
  \xi^L_n = g_n(\phi)\partial_{\phi} -rg'_n(\phi)\partial_r.
\end{align}
The naive procedure using
\eqref{eq:asymptotic_charge_Q_formula},
\eqref{eq:asymptotic_charge_k_formula},
\eqref{eq:asymptotic_charge_ktilde_formula},
\eqref{eq:Poisson_bracket_formula} and \eqref{eq:central_charge_formula},
together with
\begin{align}
  f_n(t) = -\frac{\beta}{2\pi}e^{-\frac{2\pi}{\beta}int},
\qquad
  g_n(\phi) = -e^{-in\phi},
\end{align}
gives the values of $c_{R,L}$ as
\begin{align}
  c_R=0,\qquad c_L=\frac{3k}{G_4}\int\!d\theta\,A(\theta)B(\theta)F(\theta),
\end{align}
in a similar way to \S\ref{sec:AdS3_naive_KerrCFT}.
If we expected that a non-chiral CFT$_2$ is dual to this geometry with an appropriate boundary condition,
the vanishing $c_R$ is not a reasonable one.
Then let us try a similar regularization procedure to \S\ref{sec:AdS3_regularization}.
We shift the orbifolding from \eqref{eq:4d_general_NHEK_naive_periodicity} to
\begin{align}
  (t',\phi')\sim (t',\phi'+2\pi),
\qquad
  \text{where}
\quad
  (t',\phi')=(t+\alpha\phi,\phi).
\label{eq:4d_general_NHEK_regularized_periodicity}
\end{align}
For the current case, the meaning of this shift is subtle,
because $\Sigma'$ is not always space-like in this case.
That is, it may also become time-like, depending on the value of $\theta$ in general.
Furthermore, it is difficult to justify this regularization as the residue from the near horizon parameter
\eqref{eq:zeroentropy_origin_of_regularization}, 
because we now have a finite $\epsilon$ in the words of Appendix \ref{sec:limits}.
That is, in the derivation of \eqref{eq:limits_MP_NH_ZE3},
we dropped the terms proportional to $\lambda$
and kept those proportional to $C=\lambda/\epsilon$ there.
It is, usually, nonsense for finite $\epsilon$,
and could be justified only as an $\epsilon$ expansion when $\epsilon$ is infinitesimally small.
We can point out a very similar problem in \S 6 of \cite{Matsuo:2010ut}.

Anyway, if we introduce the regularization, \eqref{eq:4d_general_NHEK_regularized_periodicity}
induces a natural periodicity for $t$ and we can take bases as
\begin{align}
  f_n(t) = -\alpha e^{-in\frac{t}{\alpha}},
\qquad
  g_n(\phi) = -e^{-in\phi},
\end{align}
leading to
\begin{align}
  c_R = \frac{3k}{G_4}\int\!d\theta\,A(\theta)B(\theta)F(\theta).
\label{eq:4d_general_cR}
\end{align}
This is a desirable result for us.
At the same time, 
calculation of $c_L$ for \eqref{eq:4d_general_NHEK}
under \eqref{eq:4d_general_NHEK_regularized_periodicity}
gives
\begin{align}
  c_L = 
  \lim_{\alpha\to 0,\;r\to\infty} 
  \bigg[
  \frac{52k\alpha r}{G_4}\int\!d\theta\,\frac{A(\theta)F(\theta)}{B(\theta)}\big(k^2B(\theta)^2-1\big)
  + \frac{3k}{G_4}\int\!d\theta\,A(\theta)B(\theta)F(\theta)
  \bigg].
\label{eq:4d_general_cL}
\end{align}
It yields the desirable value
\begin{align}
  c_L = \frac{3k}{G_4}\int\!d\theta\,A(\theta)B(\theta)F(\theta),
\end{align}
if $\alpha r\to 0$ in the limit.
This would be satisfied if the $\epsilon$-expansion in \eqref{eq:limits_MP_NH_ZE3}
were to be justified for $\epsilon\sim 1$,
since in that case $C\sim\lambda$ and then
$\alpha r = Cr/2\sim\lambda r$, which goes to $0$ by definition of the near horizon limit.
However, we again stress that we cannot trust the approximation of
the $\epsilon$-expansion at all in general although
there would be some unknown ways to justify it.
These observations may throw light on our exploration for
non-chiral Kerr/CFT on general extremal or non-extremal black holes.
We leave it for a future work.

To avoid confusion, we comment about \eqref{eq:4d_general_cL}
in the case of the zero entropy limit.
At a glance, the first term appears to be proportional to $Cr$
and not to vanish generically in the limit,
because
\begin{align}
  k\sim\frac{1}{\sqrt{\epsilon}},
\qquad
  B(\theta)\sim\sqrt{\epsilon},
\qquad
  (k^2B(\theta)^2 -1) \sim\epsilon,
\end{align}
as was shown in \cite{Azeyanagi:2010pw}.
Unlike $\lambda r$, there is no guarantee that $Cr$ goes to $0$.
Actually, the discussion above using \eqref{eq:4d_general_NHEK} is
not a precise one.
Now that we have revived small $\lambda$, then the geometry is also altered due to it.
Explicit calculation for the 5D extremal Myers-Perry black hole in Appendix~\ref{sec:limits}
shows that, 
when we revive the terms in the metric proportional to $\lambda$,
the extra terms in the central charges are proportional to
some positive powers of $\lambda r$, rather than $Cr$.
Therefore it gives the desirable values for the central charges again,
and we expect that it is also true for the general case  other than Myers-Perry case.

\section{On the Local Transformation from Usual AdS$_3$ to Near Horizon Extremal BTZ Geometry}
\label{sec:mapping}

In this appendix, we will describe a coordinate transformation
between the asymptotic regions of 
\eqref{eq:NHEK_AdS3} and \eqref{eq:AdS3_conventional_AdS3}.

A mapping between \eqref{eq:NHEK_AdS3} and \eqref{eq:AdS3_conventional_AdS3}
is given as follows.
First, \eqref{eq:AdS3_conventional_AdS3} is connected with
\eqref{eq:AdS3_by_wy} by
\begin{align}
  y = \frac{1}{\rho},\qquad w_{\pm} = \psi\pm\tau,
\end{align}
and \eqref{eq:AdS3_by_wy} is transformed to \eqref{eq:NHEK_AdS3}
through \eqref{eq:mapping_w}.
Then in total, the transformation is written as
\begin{align}
  \tau + \psi = e^{\phi},
\qquad
  \tau - \psi = \frac{1}{2}\Big(t+\frac{1}{r}\Big),
\qquad
  \rho^2 = re^{-\phi}.
\label{eq:transf_fromAdS_toNHEK_exact}
\end{align}

By using this transformation
\eqref{eq:transf_fromAdS_toNHEK_exact},
we can map the boundary conditions and the ASG generators
on \eqref{eq:AdS3_conventional_AdS3} to those on \eqref{eq:NHEK_AdS3}.
The most famous boundary condition is that discovered in \cite{Brown:1986nw},
\begin{align}
  h_{\mu\nu}\sim
  \begin{pmatrix}
    1 & \rho^{-3} & 1\\
      & \rho^{-4} & \rho^{-3}\\
      &       & 1
  \end{pmatrix},
  \label{eq:AdS3_BH_bc}
\end{align}
but here we adopt the recently proposed, 
loser boundary condition \cite{Porfyriadis:2010vg},
\begin{align}
  h_{\mu\nu}\sim
  \begin{pmatrix}
    1 & \rho^{-1} & 1\\
      & \rho^{-4} & \rho^{-1}\\
      &       & 1
  \end{pmatrix}.
  \label{eq:AdS3_PW_bc}
\end{align}
Then the ASG is generated by 
\begin{align}
  \xi =&
  \big[T^L(\tau+\psi) + T^R(\tau-\psi) + \mathcal{O}(\rho^{-2})\big]\partial_{\tau}
\nonumber\\
  &- \big[\rho\big({T^L}'(\tau+\psi) + {T^R}'(\tau-\psi)\big) + \mathcal{O}(\rho^{-1})\big]\partial_{\rho}
\nonumber\\
  &+ \big[T^L(\tau+\psi) - T^R(\tau-\psi) + \mathcal{O}(\rho^{-2}) \big]\partial_{\psi}.
  \label{eq:PW_ASG_form}
\end{align}
This boundary condition \eqref{eq:AdS3_PW_bc} is mapped by
\eqref{eq:transf_fromAdS_toNHEK_exact} to \eqref{eq:AdS3_KerrCFT_bc},
and the ASG generators \eqref{eq:PW_ASG_form} goes to
\begin{align}
  \xi =&
  \Big[f\big(t+\frac{1}{r}\big) -\frac{1}{r}f'\big(t+\frac{1}{r}\big)
       -\frac{1}{r}g'(\phi)
      + \mathcal{O}\big(\sfrac{1}{r}\big)
    \Big]\partial_t
\nonumber\\&
    + \Big[-f'\big(t+\frac{1}{r}\big)r 
           -g'(\phi)r
      + \mathcal{O}(1)
    \Big]\partial_r
    + \Big[g(\phi)
           + \mathcal{O}\big(\sfrac{1}{r}\big)\Big]\partial_{\phi}
\nonumber\\
     =&
  \Big[f(t)
      + \mathcal{O}\big(\sfrac{1}{r}\big)
    \Big]\partial_t
    + \Big[-f'(t)r 
           -g'(\phi)r
      + \mathcal{O}(1)
    \Big]\partial_r
    + \Big[g(\phi)
           + \mathcal{O}\big(\sfrac{1}{r}\big)\Big]\partial_{\phi},
  \label{eq:AdS3_ASG_form_fromPW}
\end{align}
where we defined $f(t)$ and $g(\phi)$ as
\begin{align}
  f(t) \equiv 4T^R(\sfrac{t}{2}),
  \qquad
  g(\phi) \equiv 2e^{-\phi}T^L(e^{\phi}).
\end{align}
This translated generators \eqref{eq:AdS3_ASG_form_fromPW}
have the same form as
\eqref{eq:AdS3_KerrCFT_ASG_form},
including the subleading orders.

\section{Different Limits for Near\,Horizon and Zero\,Entropy}
\label{sec:limits}

In this appendix, we investigate some possible ways to take
near horizon and zero entropy limits for an extremal black hole
simultaneously. 
We can see that an infinitesimally orbifolded AdS$_3$ structure emerges in any case,
but the resulting form or the orbifolding of the AdS$_3$
differs depending on the relative speed of these two limits.

As one of the simplest examples,
we take the 5d extremal Myers-Perry black hole.
We expect the following analysis would be valid in the general case.
The metric of the 5d extremal Myers-Perry black hole is given by
\begin{align}
  ds^2 &=
  -d\hat{t}^2
  + \frac{\Xi \hat{r}^2}{\Delta}d\hat{r}^2
  + \frac{(a+b)^2}{\Xi}(d\hat{t} - a\sin^2\theta d\hat{\phi} - b\cos^2\theta d\hat{\psi})^2
\nonumber\\
  &\quad
  + (\hat{r}^2 + a^2)\sin^2\theta d\hat{\phi}^2
  + (\hat{r}^2 + b^2)\cos^2\theta d\hat{\psi}^2
  + \Xi d\theta^2,
\label{eq:ext_MP}
\end{align}
where $a\ge b \ge 0$ and
\begin{align}
\Xi = \hat{r}^2 + a^2\cos^2\theta + b^2\sin^2\theta,
\qquad
\Delta = (\hat{r}^2-ab)^2.
\end{align}
The ranges of the angular coordinates $\theta$, $\hat{\phi}$, $\hat{\psi}$ are
\begin{align}
  0\le\theta\le\frac{\pi}{2}, 
\qquad
  \hat{\phi}\sim\hat{\phi} + 2\pi, 
\qquad
  \hat{\psi}\sim\hat{\psi} + 2\pi.
\end{align}
The Bekenstein-Hawking entropy $S_{\mathit{BH}}$ and the
ADM mass $M$ are given, respectively, by
\begin{align}
S_{\mathit{BH}} = \frac{\pi^2}{2G_5}(a+b)^2\sqrt{ab},
\qquad
M = \frac{1}{2G_5}(a+b)^2.
\end{align}
Therefore, the zero entropy limit, keeping the mass non-zero,
is described by
\begin{align}
  a \ne 0,\qquad b = \epsilon a, \qquad \epsilon\to 0.
  \label{eq:limits_zeroentropy}
\end{align}
On the other hand, the near horizon limit for this geometry is given by 
defining a new coordinate system \cite{Bardeen:1999px}
\begin{align}
  \hat{\phi} = \phi + \frac{1}{a+b}\hat{t}, 
\qquad
  \hat{\psi} = \psi + \frac{1}{a+b}\hat{t}, 
\qquad
  \hat{r}^2 = ab + \lambda (a+b)^2r, 
\qquad
  \hat{t}   = \frac{a}{\lambda}\,t,
  \label{eq:limits_nearhorizon_transformation}
\end{align}
and taking
\begin{align}
  \lambda \to 0.
  \label{eq:limits_nearhorizon}
\end{align}
When \eqref{eq:limits_zeroentropy} and \eqref{eq:limits_nearhorizon} supplemented by 
\eqref{eq:limits_nearhorizon_transformation} are taken at the same time, 
the metric goes to
\begin{align}
  ds^2 =\;
  &\frac{a^2}{\cos^2\theta}r^2dt^2
  + \frac{a^2\cos^2\theta}{4}\frac{dr^2}{r^2}
  + a^2\cos^2\theta d\theta^2
\nonumber\\
  &+ 2a^2(\sin^2\theta + \tan^2\theta)rdtd\phi
  + a^2\tan^2\theta d\phi^2
  + 2a^2\cos^2\theta rdtd\psi
\nonumber\\
  &+ 2a^2(1+\cos^2\theta)\epsilon rdtd\psi
  + 2a^2\sin^2\theta \epsilon d\phi d\psi
  + a^2\cos^2\theta (\epsilon + \lambda r)d\psi^2
  + \dots.
\label{eq:limits_MP_NH_ZE0}
\end{align}
where we kept the subleading terms (the third line)
up to first order of $b$ or $\lambda$, only for those including $d\psi$.
This is a regular geometry, 
but,  
in order to obtain an orbifolded AdS$_3$ structure,
we will carry out an additional
scaling transformation for $t$ and $\psi$.

In that scaling, the ratio
\begin{align}
  C\equiv\lim_{\lambda,\epsilon\to 0} \frac{\lambda}{\epsilon},
\label{eq:limits_defof_C}
\end{align}
will be important.
First let us assume $C<\infty$.
At that time, 
we can apply a scaling transformation
\begin{align}
  t = \frac{\sqrt{\epsilon}}{2}\;\tilde{t},
\qquad
  \psi = -\frac{\tilde{\psi}}{2\sqrt{\epsilon}},
\label{eq:limits_scaling1}
\end{align}
and take the limits $\lambda\to 0$, $\epsilon\to 0$.
Then the geometry becomes
\begin{align}
  ds^2
  &= \frac{a^2\cos^2\theta}{4}
  \bigg[
  \frac{dr^2}{r^2}
  - 2r d\tilde{t}d\tilde{\psi}
  + (1 + Cr) d\tilde{\psi}^2
  \bigg]
  + a^2\cos^2\theta d\theta^2
  + a^2\tan^2\theta d\phi^2,
\label{eq:limits_MP_NH_ZE3}
\end{align}
where the periodicity is given by
\begin{align}
  (\tilde{t},\tilde{\psi})\sim(\tilde{t},\tilde{\psi} + 4\pi\sqrt{\epsilon}).
\label{eq:limits_MP_NH_ZE3_periodicity}
\end{align}
Here we notice that this metric \eqref{eq:limits_MP_NH_ZE3}
has the same form as \eqref{eq:zeroentropy_regularized},
except that \eqref{eq:limits_MP_NH_ZE3} has an additional $d\phi^2$ term
since it is a 5d metric.
Therefore the terms in the bracket is locally AdS$_3$ for any value of $C$.
In particular, in the $C\to 0$ limit, it goes to a geometry containing the same structure as the 
near horizon extremal BTZ geometry,
\begin{align}
  ds^2
  &= \frac{a^2\cos^2\theta}{4}
  \bigg[
   -r^2d\tilde{t}^2
  + \frac{dr^2}{r^2}
  + (d\tilde{t} - d\tilde{\psi})^2
  \bigg]
  + a^2\cos^2\theta d\theta^2
  + a^2\tan^2\theta d\phi^2.
\label{eq:limits_MP_NH_ZE1}
\end{align}
{}From \eqref{eq:limits_defof_C}, $C\to 0$ means $\lambda\ll\epsilon$.
It corresponds to a procedure in which we first take the near horizon limit
and then take the zero entropy limit. 
It is nothing but the one adopted in
\cite{Guica:2007gm,Nakayama:2008kg,Guica:2010ej,Azeyanagi:2010pw}.

On the other hand, if $C>0$,
we can consider another scaling
\begin{align}
  t = \sqrt{\lambda}\;\tilde{t}',
\qquad
  \psi = -\frac{\tilde{\psi}'}{\sqrt{\lambda}},
\label{eq:limits_scaling2}
\end{align}
instead of \eqref{eq:limits_scaling1}.
This is essentially equivalent to \eqref{eq:limits_scaling1}
when $0<C<\infty$,
since they are connected by a finite rescaling
\begin{align}
  \tilde{t} = 2\sqrt{C}\,\tilde{t}',
\qquad
  \tilde{\psi} = \frac{2}{\sqrt{C}}\tilde{\psi}'.
\end{align}
By \eqref{eq:limits_scaling2},
the metric \eqref{eq:limits_MP_NH_ZE0} is transformed to
\begin{align}
  ds^2
  &= a^2\cos^2\theta
  \bigg[
  \frac{dr^2}{4r^2}
  - 2r d\tilde{t'}d\tilde{\psi}'
  + (C^{-1} + r) d\tilde{\psi}'^2
  \bigg]
  + a^2\cos^2\theta d\theta^2
  + a^2\tan^2\theta d\phi^2,
\nonumber\\
  &= a^2\cos^2\theta
    \Big[-\rho^2d\tau^2 + \frac{d\rho^2}{\rho^2}
         + \rho^2d\chi^2
         + C^{-1}(d\chi - d\tau)^2
    \Big]
  + a^2\cos^2\theta d\theta^2
  + a^2\tan^2\theta d\phi^2,
\label{eq:limits_MP_NH_ZE4}
\end{align}
where in the second line we adopted a further transformation
\begin{align}
  r = \rho^2,
\qquad
  \tilde{t}' = \tau,
\qquad
  \tilde{\psi}' = \chi - \tau,
\end{align}
under which the resulting periodicity is
\begin{align}
  (\tau,\chi)\sim(\tau,\chi + 2\pi\sqrt{\lambda}).
\label{eq:limits_MP_NH_ZE2_periodicity}
\end{align}
This form of the limit with finite $C$ corresponds to the one adopted
in \S 4 of \cite{Matsuo:2010ut}.
In this form, when the $C\to\infty$ limit is taken,  we have
\begin{align}
  ds^2
  &= a^2\cos^2\theta
  \Big[-\rho^2d\tau^2 + \frac{d\rho^2}{\rho^2}
       + \rho^2d\chi^2
  \Big]
  + a^2\cos^2\theta d\theta^2
  + a^2\tan^2\theta d\phi^2.
\label{eq:limits_MP_NH_ZE2}
\end{align}
This geometry contains the conventional form for the AdS$_3$
orbifolded by \eqref{eq:limits_MP_NH_ZE2_periodicity}.
This limit, implying $\epsilon\ll\lambda$,
means that we first take the zero entropy limit and
then the near horizon limit.
It corresponds to the prescription investigated in
\cite{Bardeen:1999px}.
We stress that the two orbifoldings 
\eqref{eq:limits_MP_NH_ZE3_periodicity} in \eqref{eq:limits_MP_NH_ZE1}
and \eqref{eq:limits_MP_NH_ZE2_periodicity} in \eqref{eq:limits_MP_NH_ZE2}
are physically quite different,
as we pointed out in \S\ref{sec:AdS3}.

\section{Relation to the RG Flow in BTZ Black Hole}
\label{sec:BTZ}

In \cite{Azeyanagi:2008dk}, the current authors discussed the 
Kerr/CFT for rotating D1-D5-P black strings. Especially they took
the two sets of the asymptotic Virasoro generators for the extremal BTZ black hole appearing 
as a part of the near horizon geometry for these rotating black string, 
and then investigated the RG flow of the generators down to the ``very near horizon geometry''.%
\footnote{We sometimes use this terminology to represent the near horizon geometry for the BTZ black hole.}
As a result, it is found that the single set of asymptotic Virasoro generators, which appears
when we apply the Kerr/CFT along the Kaluza-Klein circle, can be 
interpreted as a low energy remnant of the two sets of them for the whole extremal BTZ black hole. 
In other words, at least for this special setup, the Kerr/CFT is interpreted 
as the low energy limit of AdS$_3$/CFT$_2$. 
In this appendix, we revisit this analysis and show some relation
to the result obtained in this paper.

We start with the metric of an extremal BTZ black hole \cite{Banados:1992wn, Banados:1992gq},
\begin{align}
ds^2
&=
L^2\left[
-\frac{\rho^4}{\rho^2+r_+^2}d\tau^2
+ \frac{d\rho^2}{\rho^2}
+ (\rho^2 + r_{+}^2)
  \Big(d\psi - \frac{r_+^2}{\rho^2+r_+^2}d\tau\Big)^2 
\right],
\label{eq:BTZ_metric}
\end{align}
where the periodicity is imposed as 
\begin{align}
\psi\sim\psi + 2\pi.
\label{eq:BTZ_periodicity}
\end{align}
This black hole has an event horizon at $\rho=0$
and the associated Bekenstein-Hawking entropy is
\begin{align}
  S_{\mathit{BH}} = 2\pi L r_+.
\label{eq:BTZ_Sbh}
\end{align}
We note that this geometry appears in the near horizon limit for the non-rotating D1-D5-P
black strings in the form of a direct product with $\mathrm{S}^3$.
For simplicity, we will focus on this 3d part throughout this appendix.
At the infinity, $\rho\to\infty$, this geometry asymptotes
to the AdS$_3$ \eqref{eq:AdS3_conventional_AdS3}.
For this geometry, to investigate the asymptotic symmetry, 
we can impose \eqref{eq:AdS3_BH_bc} or \eqref{eq:AdS3_PW_bc}
as a consistent boundary condition.
Here we adopt \eqref{eq:AdS3_PW_bc} again,
and then the ASG is generated by \eqref{eq:PW_ASG_form}.
From the periodicity \eqref{eq:BTZ_periodicity},
the basis is spanned by the functions
\begin{align}
  T^R_n(x) = \frac{1}{2}e^{inx},
\qquad
  T^L_n(x) = \frac{1}{2}e^{inx},
\end{align}
and the corresponding ASG generators are
\begin{subequations}
\begin{align}
  \zeta^{R}_n&=\frac{1}{2}
   \Big(e^{in(\tau+\psi)}\partial_{\tau}
   - inre^{in(\tau+\psi)}\partial_{\rho}
   + e^{in(\tau+\psi)}\partial_{\psi}\Big),\\
  \zeta^{L}_n&=\frac{1}{2}
   \Big(e^{in(\tau-\psi)}\partial_{\tau}
   - inre^{in(\tau-\psi)}\partial_{\rho}
   - e^{in(\tau-\psi)}\partial_{\psi}\Big).
\end{align}
\label{eq:BTZ_PW_generators}%
\end{subequations}
Then $\{\zeta^{R}_n\}$ and $\{\zeta^{L}_n\}$ respectively generate 
two sets of Virasoro symmetry. 

With the coordinates transformation from $(\tau,\rho,\psi)$ to $(t,r,\phi)$,
\begin{align}
  \rho^2 &= \correctedA{\frac{\lambda r}{2}}, \quad \tau=-\frac{r_+}{\lambda}\,t,
\quad \psi = \phi-\frac{r_+}{\lambda}\,t, 
\label{eq:BTZ_verynear_transformation}
\end{align}
the (very) near horizon limit for this geometry is given by $\lambda\to 0$.

\correctedA{For a while, we consider the behavior of the transformation
\eqref{eq:BTZ_verynear_transformation}
without the limit of $\lambda\to 0$.
Under it,}
the metric \eqref{eq:BTZ_metric} is written as 
\begin{align}
  ds^2 &=
  L^2\left[
  \frac{dr^2}{4r^2}
  - r_{\!+}\,rdtd\phi
  + \Big(r_{\!+}^2 + \frac{\lambda}{2}\,r\Big)d\phi^2
  \right].
\label{eq:BTZ_verynear_metric}
\end{align}
At the same time, the ASG generators \eqref{eq:BTZ_PW_generators}
are transformed to
\begin{subequations}
\begin{align}
  \zeta^R_n &= -\bigg(\frac{\lambda}{2r_+}\partial_t+inr\partial_r\bigg)e^{-in(\frac{2r_+}{\lambda}t-\phi)}, 
\label{eq:BTZ_verynear_ASG_generators_right}
\\
  \zeta^L_n &= -\bigg(\frac{\lambda}{2r_+}\partial_t+inr\partial_r+\partial_\phi\bigg)e^{-in\phi}.
\label{eq:BTZ_verynear_ASG_generators_left}
\end{align}
\label{eq:BTZ_verynear_ASG_generators}%
\end{subequations}
By defining
\begin{align}
  \phi = \frac{\tilde{\phi}}{2r_{\!+}},
\end{align}
\correctedA{the metric \eqref{eq:BTZ_verynear_metric} becomes}
\begin{align}
  ds^2 &=
  \frac{L^2}{4}\left[
  \frac{dr^2}{r^2}
  - 2rdtd\tilde{\phi}
  + \Big(1 + \frac{\lambda}{2r_{\!+}^2}\,r\Big)d\tilde{\phi^2}
  \right],
\label{eq:BTZ_verynear_zeroentropy_metric}
\end{align}
and the ASG generators \eqref{eq:BTZ_verynear_ASG_generators} goes to
\begin{subequations}
\begin{align}
  \zeta^R_n &= -\bigg(\frac{\lambda}{2r_+}\partial_t+inr\partial_r\bigg)e^{-in(\frac{4r_+^2}{\lambda}t-\tilde{\phi})/(2r_+)}, \\
  \zeta^L_n &= -\bigg(\frac{\lambda}{2r_+}\partial_t+inr\partial_r+2r_+\partial_{\tilde{\phi}}\bigg)e^{-in\frac{\tilde{\phi}}{2r_+}}.
\end{align}
\label{eq:BTZ_verynear_zeroentropy_ASG_generators}%
\end{subequations}
Now we notice that 
\correctedA{\eqref{eq:BTZ_verynear_zeroentropy_metric} and}
\eqref{eq:BTZ_verynear_zeroentropy_ASG_generators}
\correctedA{have} exactly the same \correctedA{forms} as
\correctedA{\eqref{eq:AdS3regularized} and \eqref{eq:AdS3_KerrCFT_generators}, respectively,%
\footnote{
We thank G. Moutsopulos for pointing out this fact.
}}
under identifications
\begin{align}
  \alpha \leftrightarrow \frac{\lambda}{4r_+^2},
  \qquad
  \correctedA{\ell} \leftrightarrow 2r_+.
\end{align}
\correctedA{This fact justifies the regularization procedure we adopted
in \S\ref{sec:AdS3_regularization},
as a remnant of the infinitesimal near horizon parameter $\lambda$.}

\correctedA{Now let us consider the (very) near horizon limit, $\lambda\to 0$.}
When $r_+$ takes a non-zero finite value, by taking $\lambda\to 0$ limit,
the right Virasoro generators \eqref{eq:BTZ_verynear_ASG_generators_right}
oscillates infinitely fast (except for $n=0$, when $\zeta^R_0$ goes to zero)
while the left Virasoro generators has the same form as the ones appearing in the Kerr/CFT.

One may suspect that this flow of the ASG generators is not justified,
because the generators can contain some additional terms
with higher order in $\sfrac{1}{\rho}$ and they are subleading at the AdS$_3$ boundary but are dominant in the near horizon region.
However, under the transformation \eqref{eq:BTZ_verynear_transformation}, these terms diverge in $\lambda\to 0$ limit.
This will mean that the generators including these terms
change the geometry outside the very near horizon region.
At the same time, since these terms correspond to the choice of gauge
and then do not have any physical significance,
we can choose them as we like.
In order that the generators stay in the very near horizon geometry,
these terms should be chosen to be simply zero.
Therefore 
we have only to consider the leading terms \eqref{eq:BTZ_PW_generators} only.%
\footnote{
In the Kerr/CFT, we impose the constraint $\partial_t=0$,
which will be justified by the existence of the gap in 
the spectrum of $\partial_\tau$. Because of this constraint,
we can see that no terms will be non-zero finite except the leading order terms.
Note that if we consider the case in which the leading terms diverge 
like the right Virasoro generators, 
then the generators are absent in the low energy limit. 
This is because they transform the low energy modes to the high energy modes
which were integrated out.}

\correctedA{When we take the zero entropy limit $r_+\to 0$
in addition to $\lambda\to 0$,}
the generators of both two sets of Virasoro symmetries
vibrate infinitely fast, except for $n=0$. 
It indicates that all the degrees of freedom freeze out in this limit. 
This result is consistent with the fact that the entropy of the system goes to zero.
At the same time, because these ``frozen'' Virasoro generators
are derived as the ``IR limit'' for the ASG generators 
\eqref{eq:BTZ_PW_generators},
we can say that the non-chiral Kerr/CFT for the (very) near horizon geometry 
in the zero entropy limit is ``UV completed'' by 
AdS$_3$/CFT$_2$ for the extremal BTZ black hole.  
Although this observation is rather speculative and is not based on a rigid discussion,
it may help us to understand the non-chiral Kerr/CFT better.



\providecommand{\href}[2]{#2}\begingroup\raggedright\endgroup
\end{document}